\documentclass{aa}  
\usepackage{graphicx}
\usepackage{txfonts}
\def\ltapprox{\raise 2pt \hbox {$<$} \kern-1.1em \lower 5pt \hbox {$\approx$}}

\def\ltsim{\; \raise0.3ex\hbox{$<$\kern-0.75em \raise-1.1ex\hbox{$\sim$}}\; }
\def\gtsim{\; \raise0.3ex\hbox{$>$\kern-0.75em \raise-1.1ex\hbox{$\sim$}}\; }

\def\ie{{\it i.e.,~}}
\def\iee{{\it i.e.~}}
\def\eg{{\it e.g.,~}}
\def\egg{{\it e.g.~}}
\begin{document}
   \title{Revised statistics of radio halos and the re-acceleration model}


   \author{R. Cassano\inst{1,2}\fnmsep\thanks{\email{rcassano@ira.inaf.it}},G. Brunetti\inst{2}, T. Venturi\inst{2}, G. Setti\inst{1,2}, D. Dallacasa\inst{1,2}, S. Giacintucci\inst{2}, S. Bardelli\inst{3}}
\authorrunning{R. Cassano et al.}

   \offprints{R.Cassano}

   \institute{Dipartimento di Astronomia, Universit\`a di Bologna, via Ranzani 1, I-40127 Bologna, Italy\\
         \and
             INAF - Istituto di Radioastronomia, via P. Gobetti 101,I-40129 Bologna, Italy\\
         \and
	    INAF - Osservatorio Astronomico di Bologna, via Ranzani 1, 40127 Bologna, Italy }

   \date{Received 4 November 2007 ; accepted 18 December 2007}

\abstract
{} 
{The statistical properties of radio halos can be used to discriminate among the possible models for their origin. Therefore an unbiased and exhaustive investigation in this direction is crucial.}
{With this goal in mind in this paper we revise the occurrence of radio halos in the redshift range 0-0.4, combining the low redshift ($z<0.2$) statistical study of XBACs
clusters with the NVSS (by Giovannini et al. 1999) with our recent results from the radio follow up of REFLEX and eBCS clusters, the GMRT radio halo survey, at higher redshift ($0.2<z<0.4$).}
{We find a significant statistical evidence (at $3.7\sigma$) of an increase of the fraction of clusters with Radio Halos with the X-ray luminosity (mass) of the parent clusters and show that this increase is in line with statistical calculations based 
on the re-acceleration scenario. 
We argue that a fundamental expectation of this scenario is  
that the probability to have radio halos emitting at
hundred MHz is larger than that at GHz frequencies and thus that future radio interferometers operating at low frequencies, such as LOFAR and LWA, should detect a 
larger number of radio halos with respect to that caught by present GHz observations.
We also show that the expected increase of the fraction of clusters with radio halos with the cluster mass as measured with future LOFAR and LWA surveys should be less
strong than that in present surveys.}
 {}

\keywords{Radiation mechanism: non--thermal - galaxies: clusters: general - 
radio continuum: general - X--rays: general}

\maketitle

\section{Introduction}

Radio halos (RH elsewhere) are diffuse Mpc--scale radio sources
observed at the center of a fraction of massive galaxy clusters 
(\egg Feretti 2005 for a review).
These sources are synchrotron emission from GeV electrons diffusing through
$\mu$G magnetic fields and provide the most important evidence of non 
thermal components in the ICM.

The clusters hosting RHs are always characterized by a peculiar dynamical 
status indicative of very recent or ongoing merger events 
(\eg Buote 2001; Schuecker et al 2001) and this suggests a connection
between the mergers and the origin of non thermal components in the ICM.
The main difficulty in understanding the origin of the synchrotron
emitting electrons in RHs is that the life--time of these electrons is
much shorter than the diffusion time necessary to cover the scale
of RHs.
Two main possibilities have been explored to explain RHs:
{\it i)} the {\it secondary electron} models, 
whereby the relativistic electrons are secondary products of the hadronic 
interactions of cosmic rays (CR) with the ICM 
(\eg Dennison 1980; Blasi \& Colafrancesco 1999), and {\it ii)}
the so-called {\it re-acceleration} models, whereby
relativistic electrons injected in the intra cluster medium (ICM) are
re-energized {\it in situ} by various mechanisms associated with the
turbulence generated by massive merger events (\eg Brunetti et al. 2001;
Petrosian et al. 2001).

The two scenarios have very different expectations for the 
basic statistical properties of RHs. 
In particular in the {\it secondary electron} models RHs should be 
very long--living phenomena.
Since the CR proton component is expected to be accumulated
in clusters, the radio emission from secondary particles
should be dominated at any time by the pile up of CRs protons during the merger 
history of the cluster, rather than by the latest merger event.
On the contrary, in the {\it re-acceleration} models RHs should 
be {\it transient phenomena} with a relatively short life--time 
(1 Gyr or less) because of the finite dissipation time-scale 
of the turbulence.
The recent discovery that galaxy clusters with similar X--ray
luminosity have a clear bi--modal distribution in 
radio properties with only a fraction of these clusters
hosting a RH (Brunetti et al. 2007) is in line with this latter
scenario.

Particle re--acceleration also provides a natural way to explain 
the morphology and spectral properties of 
the few well studied RHs and to account for the 
link between RHs and cluster mergers (e.g. Petrosian 2003;
Brunetti 2004; Blasi 2004; Feretti 2005 for reviews).
The physics of collisionless turbulence and the process of stochastic
particle re-acceleration is complex and rather poorly understood, however recent calculations have shown that there may be room for sufficient Alfv\'enic and magnetosonic
acceleration of particles in the ICM during cluster mergers
(Brunetti et al.~2004; Brunetti \& Lazarian 2007) thus providing some
physical support to this scenario.

Recent calculations in the framework of the {\it re-acceleration} scenario
have modelled the connection between RHs and cluster mergers in a 
cosmological framework, and derived the expected fraction of clusters with
RHs as a function of the mass, redshift and dynamical status of the clusters 
(Cassano \& Brunetti 2005, CB05; Cassano, Brunetti \& Setti 2006, CBS06).
These calculations provide a unique predictive power and allow 
deep radio observations of samples of galaxy clusters to test the
scenario and to constrain model parameters.

Observational studies of the statistical properties 
of RHs can be very effective in constraining the origin of RHs and, in
general, of the non thermal components in galaxy clusters.
Giovannini et al. (1999, GTF99) derived the occurrence of RHs 
in the XBACs sample by inspection of the NVSS (NRAO VLA Sky Survey, 
Condon et al. 1998) at 1.4 GHz; Kempner \& Sarazin (2001) carried out 
a similar study on the ACO sample by means of the WENSS (Westerbork Northern Sky Survey, Rengelink et al. 1997) at 327 MHz. Both cluster samples are complete up to $z=0.2$.
These studies show that RHs are rare at the detection
level of the radio surveys used and that their detection rate 
increases with increasing the X-ray luminosity
of the parent clusters: 30-35\% for clusters with X-ray luminosity 
larger than $10^{45}$ $h_{50}^{-1}$ erg/s were found to host
RHs (GTF99).

Despite these observational claims 
it is not clear whether the rarity of RHs and
the increase of their occurrence with cluster mass is real or
driven by selection biases due to the brightness limit
of the NVSS and WENSS surveys (\eg Rudnick et al. 2006).

In this paper we extend these analysis up to larger redshift, 
z$\ltsim0.4$.
Most important, we derive an unbiased statistics of RHs as we 
discuss the relevant radio and X--ray selection effects
and limit our analysis to a cluster sub--sample which is not
affected by these effects.
The starting sample used is this paper
is the combination of the NVSS--XBAC sample at $z<0.2$
and the recent sample of 50 massive (X-ray luminous)
clusters at $0.2\leq z\leq 0.4$ (Venturi et al. 2007)
observed at 610 MHz with the Giant Metrewave
Radio Telescope (GMRT).
The sample is presented in Sect.2 and the selection effects are 
discussed in Sect.3. In Sect.4 we give the results on the occurrence
of RHs and in Sect.5 they are compared with expectations
from the re--acceleration scenario. Finally, in Sect.6 we discuss the 
main implications of this model for future low frequency radio surveys.

A $\Lambda$CDM 
($H_{o}=70$ Km $s^{-1}$ $Mpc^{-1}$, $\Omega_{m}=0.3$, $\Omega_{\Lambda}=0.7$) 
cosmology is adopted.

\section{Selection of the cluster samples}

In this Section we describe the sample of galaxy clusters
in the redshift bin $0-0.4$ from which we start our statistical 
analysis for the presence of RH. It consists of two X--ray selected cluster samples, at low and at high redshifts, with already available radio follow ups for the search of RH. 
Being X-ray selected the sample may contain cooling core (CC) clusters. 
Because all RH are found in merging clusters it is well known that a possible anti-correlation exists between the presence of CCs and of giant RHs at the cluster center (Edge et al. 1992; Feretti 2000).
However, since our main aim is to perform an unbiased analysis of the occurrence of
RH in galaxy clusters independently of their dynamical status, we will consider
all clusters regardless of the presence of CCs.

\subsection{The sample at z$\leq0.2$}
Following GTF99 at low redshifts, $z\leq0.2$, 
we use the X-ray-brightest Abell-type clusters sample (XBACs, Ebeling et al.
1996).
The XBACs is a complete, all sky X-ray flux limited sample of  
242 clusters which are optically selected from the catalogs of Abell (1958)
and Abell, Corwin \& Olowin (1989, hearafter ACO) and 
compiled from the {\it ROSAT} All-Sky Survey. The sample is 
limited to high galactic latitude ($\mid b\mid\geq20^\circ$) and 
is statistically complete for X-ray fluxes larger than $5.0\times 10^{-12} 
\mathrm{erg s^{-1} cm^{-2}}$ in the 0.1-2.4 keV band up to a redshift $0.2$,
the nominal completeness limit of the ACO clusters (Ebeling et al.~1996),
but there are also 24 clusters at redshift larger than 0.2 which meet the flux 
criterion.

This sample was cross-checked with the NVSS to search for RHs (GTF99).
The NVSS is a radio survey performed at 1.4 GHz with the Very 
Large Array (VLA) in configurations $D$ and $DnC$.
It has an angular resolution of $45''$ (HPBW), a surface
brightness r.m.s. $=$0.45 mJy/beam (1 $\sigma$) and covers all the 
sky north of $\delta=-40^{\circ}$. 
Given the short baseline sampling available at the VLA, the NVSS is insensitive 
to diffuse sources on scales $\geq 15'$, and this precludes the possibility to detect RH at $z<0.044$ (assuming $1$ Mpc size, GTF99).
Thus as a starting low redshift sub--sample we extract 
from the XBACs sample all clusters with $0.044\leq z \leq0.2$ and 
with $\delta>-40$. The final sample consists of
182 clusters (excluding A\,1773 
and A\,388 which fall in 
the few remaining gaps of the NVSS).

\subsection{The sample in the redshift range: $0.2<z<0.4$}

The sample at larger redshift, $0.2<z<0.4$, consists of the combination of
two X--ray sub-samples with a relatively deep radio follow up to search for RH 
recently performed at 610 MHz with the GMRT.
The two X--ray samples are the ROSAT--ESO Flux Limited 
X--ray (REFLEX) galaxy cluster catalog (B\"oringher et al. 2004) and 
the extended ROSAT Brightest Cluster Sample (eBCS) catalog 
(Ebeling et al. 1998, 2000). 
These two catalogs have almost the same flux limit in the $0.1-2.4$ keV 
band ($\gtsim 3\cdot 10^{-12} \rm{erg\,s^{-1}\,cm^{-2}}$) and their
combination yields a homogeneous flux limited sample of clusters. 
GMRT radio follow up of these catalogues has been performed by our
group only for clusters with L$_{\rm X}$(0.1--2.4 keV) $\geq$ 
5 $\cdot$ 10$^{44}$ erg s$^{-1}$
extracted from the catalogues mentioned above.
Description of the two resulting radio--X-ray samples is given below.

\subsubsection{The REFLEX sub-sample}

The REFLEX survey covers the southern sky up to a declination 
$\delta=+2.5^{\circ}$, avoiding the Milky Way and the regions 
of the Magellanic clouds, for a total area of $13924\,\mathrm{deg}^2$ 
($4.24$ sr). The sample is complete for X-ray fluxes larger than 
$\sim3\cdot 10^{-12}\rm{erg\,s^{-1}\, cm^{-2}}$ up to $z\sim 0.3$;
above this redshift only very luminous objects (with X-ray luminosities 
of several $10^{45}$ erg/s) have been observed (B\"oringher et al. 2001).
In order to have a good u-v coverage for the radio observations at the GMRT 
in this sample we selected only clusters with $\delta\geq-30^{\circ}$ (Venturi et al. 2007). 

\begin{figure}
\begin{center}
\includegraphics[width=0.45\textwidth]{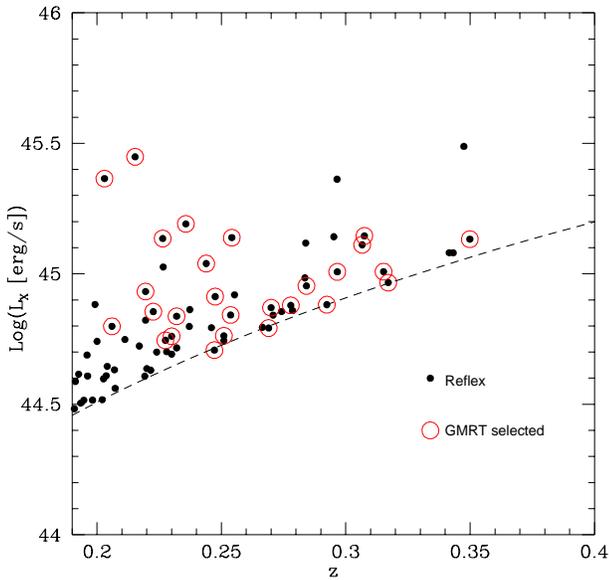}
\caption[]{X-ray luminosity ($0.1-2.4$ keV) versus $z$ for the REFLEX 
clusters (black filled circles). Open red circles indicate the clusters 
belonging to our sample.}
\label{Cap7.fig.reflex_GMRT}
\end{center}
\end{figure} 

In Fig.~\ref{Cap7.fig.reflex_GMRT} we report the distribution of 
the REFLEX clusters in the plane $L_X-z$ and with red 
circles we highlight the clusters matching our luminosity and declination criteria. 
We obtain a total sample of 27 clusters. 
The source list is reported in Tab.~\ref{Cap7.tab.reflex},
where we give (1) the REFLEX name, (2) alternative name from other catalogs, 
(3) and (4) J2000 coordinates, (5) redshift, (6) the X--ray luminosity in the 
0.1--2.4 keV band in unit of $10^{44}$ erg $s^{-1}$, (7) information on the diffuse emission. 

%
%
\begin{table*}
\caption[]{Cluster sample from the REFLEX catalog.}
\begin{center}
\label{Cap7.tab.reflex}
\begin{tabular}{rrccccc}
\hline\noalign{\smallskip}
REFLEX Name   & Alt. name & RA$_{J2000}$ &  DEC$_{J2000}$ & z &  $L_{X}$ & diffuse emission\\
              &           &              &                &   &  $10^{44} erg/s$        & \\
\noalign{\smallskip}
\hline\noalign{\smallskip}
$^{\surd}$ RXCJ\,0003.1$-$0605 & A\,2697 &  00 03 11.8 &  $-$06 05 10 & 0.2320 &   6.876 & none\\
$^{\star}$ RXCJ\,0014.3$-$3023 & A\,2744 &  00 14 18.8 &  $-$30 23 00 & 0.3066 &  12.916 & {\bf RH}\\
$^{\surd}$ RXCJ\,0043.4$-$2037 & A\,2813 &  00 43 24.4 &  $-$20 37 17 & 0.2924 &   7.615 & none\\
$^{\surd}$ RXCJ\,0105.5$-$2439 & A\,141  &  01 05 34.8 &  $-$24 39 17 & 0.2300 &   5.762 & none\\ 
$^{\surd}$ RXCJ\,0118.1$-$2658 & A\,2895 &  01 18 11.1 &  $-$26 58 23 & 0.2275 &   5.559 & none\\ 
$^{\surd}$ RXCJ\,0131.8$-$1336 & A\,209  &  01 31 53.0 &  $-$13 36 34 & 0.2060 &   6.289 &{\bf RH}\\
$^{\surd}$ RXCJ\,0307.0$-$2840 & A\,3088 &  03 07 04.1 &  $-$28 40 14 & 0.2537 &   6.953 &none\\
$^{\circ}$ RXCJ\,0437.1$+$0043 &  $-$    &  04 37 10.1 &  $+$00 43 38 & 0.2842 &   8.989 & none\\
$^{\surd}$ RXCJ\,0454.1$-$1014 & A\,521  &  04 54 09.1 &  $-$10 14 19 & 0.2475 &   8.178 &relic\\
           RXCJ\,0510.7$-$0801 &  $-$    &  05 10 44.7 &  $-$08 01 06 & 0.2195 &   8.551 &?\\
$^{\surd}$ RXCJ\,1023.8$-$2715 & A\,3444 &  10 23 50.8 &  $-$27 15 31 & 0.2542 &  13.760 &core halo\\
$^{\surd}$ RXCJ\,1115.8$+$0129 &  $-$    &  11 15 54.0 &  $+$01 29 44 & 0.3499 &  13.579 &none\\
$^{\star}$ RXCJ\,1131.9$-$1955 & A\,1300 &  11 31 56.3 &  $-$19 55 37 & 0.3075 &  13.968 &{\bf RH}\\
           RXCJ\,1212.3$-$1816 &  $-$    &  12 12 18.9 &  $-$18 16 43 & 0.2690 &   6.197 &?\\
$^{\surd}$ RXCJ\,1314.4$-$2515 &  $-$    &  13 14 28.0 &  $-$25 15 41 & 0.2439 &  10.943 &2 relics, 1 small RH\\
$^{\surd}$ RXCJ\,1459.4$-$1811 & S\,780  &  14 59 29.3 &  $-$18 11 13 & 0.2357 &  15.531 &none\\
           RXCJ\,1504.1$-$0248 &  $-$    &  15 04 07.7 &  $-$02 48 18 & 0.2153 &  28.073 &?\\
$^{\surd}$ RXCJ\,1512.2$-$2254 &  $-$    &  15 12 12.6 &  $-$22 54 59 & 0.3152 &  10.186 &none\\
           RXCJ\,1514.9$-$1523 &  $-$    &  15 14 58.0 &  $-$15 23 10 & 0.2226 &   7.160 &?\\
$^{\star}$ RXCJ\,1615.7$-$0608 & A\,2163 &  16 15 46.9 &  $-$06 08 45 & 0.2030 &  23.170 &{\bf RH}\\
$^{\surd}$ RXCJ\,2003.5$-$2323 &  $-$    &  20 03 30.4 &  $-$23 23 05 & 0.3171 &   9.248 &{\bf RH}\\
           RXCJ\,2211.7$-$0350 &  $-$    &  22 11 43.4 &  $-$03 50 07 & 0.2700 &   7.418 &?\\
$^{\surd}$ RXCJ\,2248.5$-$1606 & A\,2485 &  22 48 32.9 &  $-$16 06 23 & 0.2472 &   5.100 &none\\ 
$^{\surd}$ RXCJ\,2308.3$-$0211 & A\,2537 &  23 08 23.2 &  $-$02 11 31 & 0.2966 &  10.174 &none\\
$^{\surd}$ RXCJ\,2337.6+0016   & A\,2631 &  23 37 40.6 &  $+$00 16 36 & 0.2779 &   7.571 &none\\
$^{\surd}$ RXCJ\,2341.2$-$0901 & A\,2645 &  23 41 16.8 &  $-$09 01 39 & 0.2510 &   5.789 &none\\ 
$^{\surd}$ RXCJ\,2351.6$-$2605 & A\,2667 &  23 51 40.7 &  $-$26 05 01 & 0.2264 &  13.651 &none\\
\noalign{\smallskip}
\hline
\end{tabular}
\end{center}
Symbols are as follows: $^{\surd}$ marks the clusters observed by us with the 
GMRT as part of our radio halo survey (venturi et al. 2007, Venturi et al. in prep);  
$^{\star}$ marks the clusters with extended radio emission known from the literature 
(A\,2744 Govoni et al. 2001; A\,1300 Reid et al. 1999; 
A\,2163 Herbig \& Birkinshaw 1994 and Feretti et al. 2001). $\circ$ marks the clusters
without extended radio emission known from the literature (RXCJ0437.1+0043 Feretti et al. 2005). The unmarked 5 clusters are part of the GMRT cluster 
Key Project (P.I. Kulkarni). 
\end{table*}

Among these 27 clusters, three are clusters known to host RHs,
\ie A\,2744, A\,1300 and A\,2163.
From the remaining 24 clusters in Tab.~\ref{Cap7.tab.reflex} we selected 
all clusters with no radio information available in the literature and 
we also excluded those belonging to the GMRT Cluster Key 
Project (P.I. Kulkarni), 
and the remaining 18 clusters (marked with the symbol $\surd$ in 
Tab.~\ref{Cap7.tab.reflex}) were observed with the GMRT 
(in several observational run from January 2005 to August 2006). 
The total REFLEX sample with radio follow up thus consists of
20 galaxy clusters (here we exclude from this sample
A\,2163 and A\,209 which are already included in the XBACs 
sample, and the clusters: RXCJ\,0510.7 $-$ 0801, RXCJ\,1212.3 $-$ 1816, 
RXCJ\,1504.1 $-$ 0248, RXCJ\,1514.9 $-$ 1523 and RXCJ\,2211.7 $-$ 0350 
which belong to the GMRT Cluster Key Project and 
still have no published data\footnote{The exclusion of these clusters from our sample does not introduce any bias because they are randomly distributed in redshift and X-ray luminosity among the clusters in the sample}).

\subsubsection{The extended BCS sub-sample}

The {\it ROSAT} Brightest Cluster Sample (BCS; Ebeling et al. 1998) is a 90 
per cent flux-complete sample of the 201 clusters of galaxies in 
the northen hemisphere selected from the {\it ROSAT} All-Sky Survey (RASS).
All these clusters have flux higher than 
$4.4\times 10^{-12}\,\mathrm{ erg\,cm^{-2}\,s^{-1}}$ in the 0.1-2.4 keV band. 
This sample is combined with a low-flux extension of the BCS 
(Ebeling et al. 2000) 
which consists of 99 clusters of galaxies with flux higher than 
$2.8\times 10^{-12}\,\mathrm{erg\,cm^{-2}\,s^{-1}}$ in the 0.1-2.4 keV band.
The combination of these two samples forms the homogeneously selected 
extended BCS (eBCS) which is statistically complete up to a redshift 
$z\sim 0.3$ (Ebeling et al. 1998, 2000). 

\begin{figure}
\begin{center}
\includegraphics[width=0.45\textwidth]{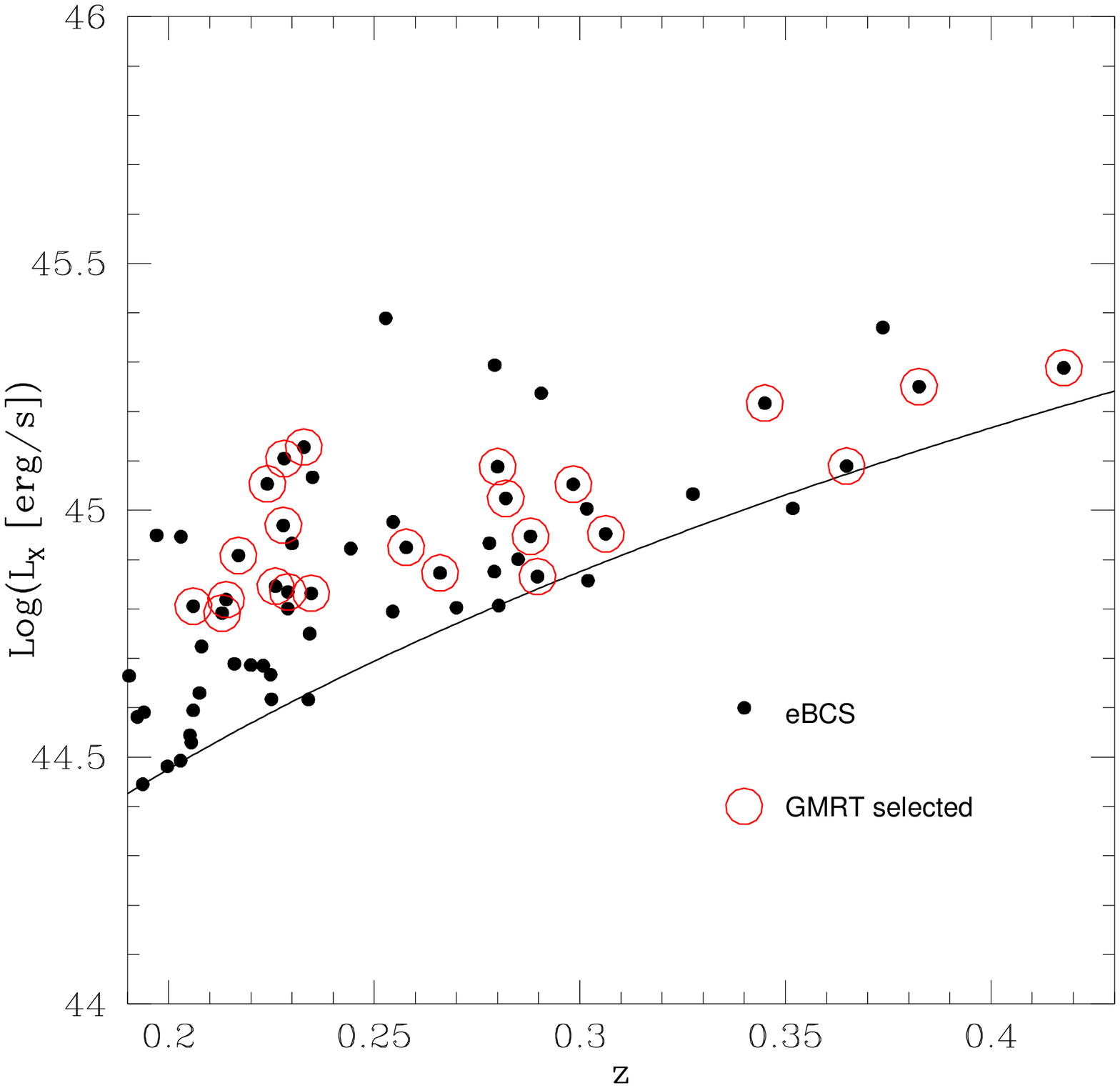}
\caption[]{X-ray luminosity ($0.1-2.4$ keV) versus $z$ for the REFLEX 
clusters (black filled circles). Open red circles indicate the clusters 
belonging to our sample.}
\label{Cap7.fig.eBCS_GMRT}
\end{center}
\end{figure}

From the eBCS catalog we selected all clusters 
with $15^{\circ}<\delta<60^{\circ}$, and obtained a total sample 
of 23 clusters (Venturi et al, in prep).
In Fig.~\ref{Cap7.fig.eBCS_GMRT} we report the distribution of the eBCS 
clusters in the plane $L_X-z$ and with red circles we highlight 
the clusters which meet our selection criteria\footnote{Formally the eBCS sub-sample 
contains also RXCJ2228.6+2037 which is at $z>0.4$ and is not included 
in the following analysis.}.

\begin{table*}
\caption[]{Cluster sample from the eBCS catalog at $z<0.4$.}
\begin{center}
\label{Cap7.tab.eBCS}
\begin{tabular}{lrcccc}
\hline\noalign{\smallskip}
Name   &  RA$_{J2000}$ &  DEC$_{J2000}$ & z &  $L_{X}$ & diffuse emission\\ 
       &               &                &   & $10^{44} erg/s$   & \\
\noalign{\smallskip}
\hline\noalign{\smallskip}
$^{\surd}$RXJ0027.6$+$2616  &  00 27 49.8  & +26 16 26 & 0.3649 & 12.29 & none \\
$^{\surd}$A611              &  08 00 58.1  & +36 04 41 & 0.2880 & 8.855 & none\\
$^{\surd}$A697              &  08 42 53.3  & +36 20 12 & 0.2820 & 10.57 & {\bf RH} \\  
$^{\surd}$Z2089             &  09 00 45.9  & +20 55 13 & 0.2347 & 6.79  & none\\
$^{\star}$A773              &  09 17 59.4  & +51 42 23 & 0.2170 & 8.097 & {\bf RH}\\ 
$^{\surd}$A781              &  09 20 23.2  & +30 26 15 & 0.2984 & 11.29 & relic ?\\
$^{\surd}$Z2701             &  09 52 55.3  & +51 52 52 & 0.2140 & 6.59  & none\\
$^{\surd}$Z2661             &  09 49 57.0  & +17 08 58 & 0.3825 & 17.79 & complex$^1$\\
$^{\surd}$A963              &  10 17 09.6  & +39 01 00 & 0.2060 & 6.39  & none\\
$^{\surd}$A1423             &  11 57 22.5  & +33 39 18 & 0.2130 & 6.19  & none\\
$^{\surd}$Z5699             &  13 06 00.4  & +26 30 58 & 0.3063 & 8.96  & none\\
$^{\surd}$A1682              &  13 06 49.7  & +46 32 59 & 0.2260 & 7.017 & complex $^1$\\ 
$^{\surd}$Z5768             &  13 11 31.5  & +22 00 05 & 0.2660 & 7.465 & none\\
$^{\star}$A1758a            &  13 32 32.1  & +50 30 37 & 0.2800 & 12.26 & {\bf RH}\\ 
A1763                       &  13 35 17.2  & +40 59 58 & 0.2279 & 9.32  & none\\
$^{\surd}$Z7160             &  14 57 15.2  & +22 20 30 & 0.2578 & 8.411 & mini halo\\
$^{\surd}$Z7215             &  15 01 23.2  & +42 21 06 & 0.2897 & 7.34  & none\\
$^{\surd}$RXJ1532.9+3021    &  15 32 54.2  & +30 21 11 & 0.3450 & 16.485 & none\\
A2111                       &  15 39 38.3  & +34 24 21 & 0.2290 & 6.83   & none\\
$^{\star}$A2219             &  16 40 21.1  & +46 41 16 & 0.2281 & 12.73  & {\bf RH}\\ 
A2261                       &  17 22 28.3  & +32 09 13 & 0.2240 & 11.31  & uncertain $^2$\\
$^{\circ}$A2390              &  21 53 34.6  & +17 40 11 & 0.2329 & 13.43  & mini halo\\
\noalign{\smallskip}
\hline  
\end{tabular}
\end{center}
Symbols are as follows: $^{\surd}$ marks the clusters observed with the 
GMRT as part of our radio halo survey (Venturi et al. 2007, Venturi et al. in prep.);  
$^{\star}$ marks the clusters with a RH known from the literature 
(A\,773 Govoni et al. 2001; A\,1758 Giovannini et al. 2006; 
A\,2219 Bacchi et al. 2003); $^{\circ}$ marks the clusters with ``mini-halo''
known from the literature (A\,2390 Bacchi et al. 2003). For clusters A~2111, A~1763 
and A~2261 we analyzed 1.4 GHz VLA archive data (Venturi et al. in prep).  
$^1$ The presence of extended radio galaxies, coupled with positive residuals at the cluster centre, does not allow to establish the presence of possible cluster diffuse emission (Venturi et al. in prep.).
$^2$ The presence of a RH at the cluster center is uncertain: only B and D-array 
VLA data are available and extended sources at the cluster center do not allow a firm conclusion (see discussion in Venturi et al. in prep.).
\end{table*}


Tab.~\ref{Cap7.tab.eBCS} gives details of the clusters extracted from
the eBCS catalogue: 
(1) the cluster name, (2) and 
(3) J2000 coordinates, (4) redshift, (5) the X--ray luminosity in the 
0.1--2.4 keV band in unit of $10^{44}$ erg/s, (6) information on the diffuse emission. 
Among these 22 clusters ($z<0.4$) 4 have already known diffuse radio emission (A\,773,
A\,1758, A\,2219, A\,2390), 15 were observed by our group
at the GMRT (September 2005-August 2006), and for the remaining  
3 clusters (A\,2111, A\,1763 and A\,2261) we analyzed 1.4 GHz data from archive 
deep pointed observations at the VLA (Venturi et al. in prep.).

The final X--ray selected sample with a radio follow up at $z<0.4$ is
thus made of 21 clusters (we also exclude A\,963 which is 
already included in the XBACs sample).

\section{Radio Selection effects} 
\label{selection_NVSS}

The final sample of X--ray selected clusters with a
radio follow up from which we start our analysis is made by the
combination of the XBAC--NVSS, the REFLEX--GMRT
and the eBCS--GMRT sub--samples and consists of 220 galaxy clusters 
in the redshift interval $0.044<z<0.4$. 
The distribution of the clusters in the $L_x-z$ plane is
reported in Fig.\ref{Fig.totsample}, where we also highlight with 
red open circles the clusters with giant RH (GTF99, Venturi et al. 2007, 
Venturi et al. in prep.).

In order to perform a statistical census of the fraction of clusters 
with giant RHs in different ranges of X--ray luminosity and redshift
it is crucial to study the selection effects on the detection of RHs.
These biases come from the strategy adopted in the radio follow up
of the different X--ray sub--samples.

In the following we discuss the selection effects on the NVSS and GMRT
cluster samples and derive a radio -- X-ray sub--sample of clusters
suitable for an unbiased statistical study of RHs.

\subsection{The XBAC-NVSS sub-sample}

\begin{figure*}
\begin{center}
\includegraphics[width=10cm,height=10cm]{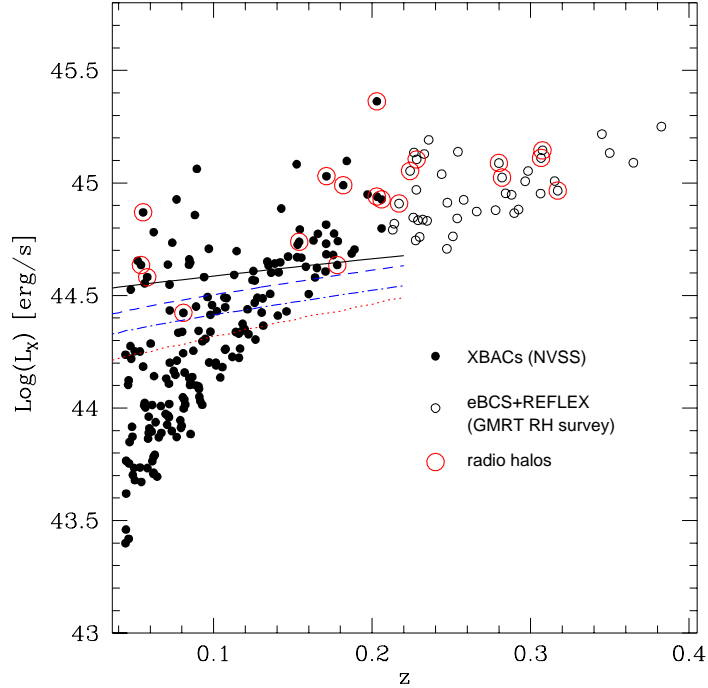}
\caption[]{X-ray luminosity ($0.1-2.4$ keV) versus $z$ for the total sample of clusters
(XBACs+REFLEX+eBCS). Open red circles indicate the clusters with known 
giant RHs. The lines give the lower limit to the cluster X-ray luminosities in the case a) (black line), in the case b) (dotted line), and in the case c) (blue dot-dashed line: 2$\sigma$, blue dashed line: 3$\sigma$) as described in the text.}
\label{Fig.totsample}
\end{center}
\end{figure*}

As previously reported, the NVSS is surface 
brightness-limited and RHs with a lower brightness are lost.

There is a clear evidence of the existence of a 
correlation between the synchrotron radio 
power at 1.4 GHz ($P_{1.4}$) of halos and the X-ray luminosity 
($L_X$) of the parent clusters 
(Liang et al. 2000; Feretti 2000, 2003; En\ss lin and R\"ottgering 2002; 
Cassano et al. 2006). This correlation is not affected by selection effects, 
at least for high X-ray luminosities (Log$(L_X)> 44.6$ erg/s;\eg Clarke 2005,
Dolag 2006, Brunetti et al. 2007) and it implies that the average brightness of 
RHs increases with the X-ray luminosity of the parent cluster
(Feretti 2005; CBS06).
If we scale the radio brightness with the cluster X-ray luminosity, 
the sensitivity limit in radio surveys will set the minimum 
surface brightness of a RH to be detected, and then a limiting X-ray luminosity of the hosting clusters.

An additional argument that should be used to discuss the selection of RHs
in brightness limited surveys is their brightness distribution.
In general, RHs are characterized by brightness profiles which smoothly decrease 
with distance from the cluster center (\eg Govoni et al. 2001). 
The outermost low brightness regions of RHs are lost in shallow surveys,
however what is important here is the capability of these surveys to detect
the central, brightest part of the diffuse emission.
Indeed the detection of these regions allows to claim the presence of
diffuse emission from clusters and to select samples of RH candidates
for a deeper radio follow up.
By making use of several well studied RHs, it was shown (Brunetti et al. 2007)
that their integrated brightness profiles are quite similar, provided that 
their radial distance is normalised to the size of the different
RHs. In particular it has been shown that about half of the total radio flux
density measured in RHs is contained in about half radius and this information
can be used in discussing the selection of RHs in brightness limited surveys.

Thus in discussing selection biases from the NVSS we use both the $P_{1.4}$--$L_X$ correlation and the constraints from the RH brightness profiles, 
and consider three possible approaches:

\begin{itemize}

\item{\it case a)}: we consider the case of RHs with a fixed radius, $R_H=500$ kpc. 
By making use of the $P_{1.4}$--$L_X$ correlation, for  
each redshift we calculate the minimum X-ray luminosity of a cluster 
that can host RHs with an expected average radio brightness of 0.45 mJy/beam within 
half radius. 
The resulting minimum X-ray luminosity 
is reported as a function of redshift
in Fig.\ref{Fig.totsample} (solid black line).

\begin{figure*}
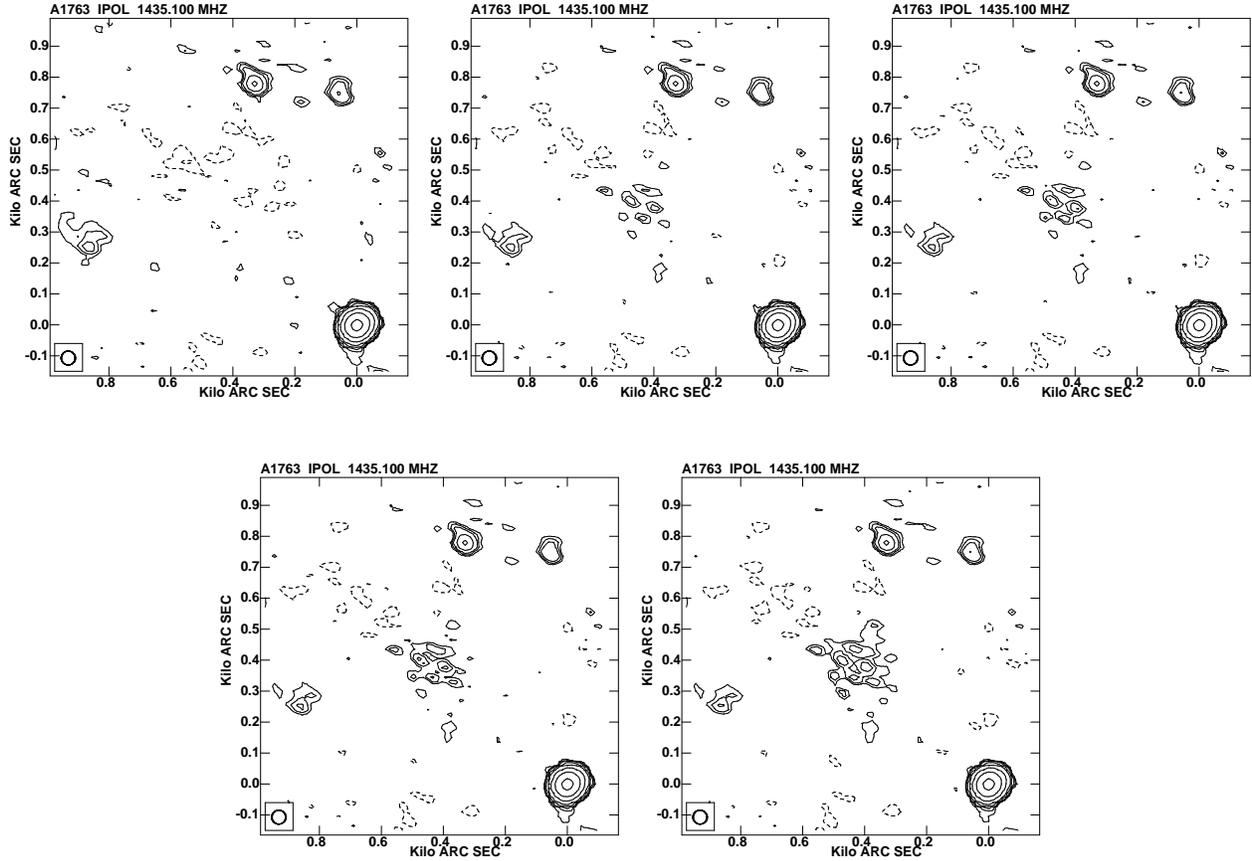

\begin{center}
\includegraphics[width=0.3\textwidth]{ROS_NO.PS}
\includegraphics[width=0.3\textwidth]{ROS_N2.PS}
\includegraphics[width=0.3\textwidth]{ROS_N3.PS}
\includegraphics[width=0.3\textwidth]{ROS_N4.PS}
\includegraphics[width=0.3\textwidth]{ROS_N6.PS}
\caption[]{Example of injection of fake RHs. We assume a size of $500$ kpc and
a flux of 0, 28, 32, 36 and 45 mJy (from top left to bottom right panel).
The r.m.s. in the images is of 0.45 mJy/beam, the HPWB is $45^{\prime\prime}\times 45^{\prime\prime}$, contours are $1.0\times (-1,1,1.4,2,4,8,32,125,500)$ mJy/beam.}
\label{Fig.fakeRH}
\end{center}
\end{figure*}

\item{\it case b)}: we consider 
an additional correlation which has been recently found between 
the size of RHs and the synchrotron radio power at 1.4 GHz, 
$P_{1.4}\propto R_H^{4.18\pm 0.68}$ (Cassano et al. 2007, C07 hearafter):
clusters with low X-ray luminosities should host 
smaller RHs than those with higher X-ray luminosities. This makes the detection of
RHs in clusters with lower X--ray luminosity easier with respect to the
previous case a).
For each $z$, we compute the minimum radio power of RHs with an average radio brightness within half radius 
(which depends on the radio power) of 0.45 mJy/beam.
The corresponding minimum X-ray luminosity of the clusters which may host
these RHs is reported in Fig.\ref{Fig.totsample}
as a function of redshift (red dotted curve).

\item {\it case c)}: we consider RHs with fixed size, $R_H=500$ kpc,
and with a typical brightness profile taken from Brunetti et al. (2007). 
For each redshift, we integrate the average brightness over an area 
as large as 4 times the NVSS beam and calculate the corresponding 1.4 GHz power 
of RHs whose brightness equals 2 and 3 times 
the rms of the NVSS (0.9 and 1.35 mJy/beam). 

\noindent
The corresponding minimum X--ray luminosity (from the $P_{1.4}-L_X$ 
correlation) of clusters which may host these RHs is reported in 
Fig.~\ref{Fig.totsample} as a function of redshift 
(blue dot-dashed and blue dashed lines mark the case of 2 and 3 times
the NVSS--rms, respectively).

\end{itemize}

In all the cases the NVSS is efficient in detecting
RHs in clusters with Log$(L_X)> 44.6$ erg/s.

In order to strengthen the reliability of the above considerations on the
efficiency of the NVSS in catching RHs, we use a more direct approach.
Adopting the same procedure of Brunetti et al. (2007) we choose a cluster without RH
and inject a ``fake'' RH in a region close to the pointing and without any significant contribution from real sources (upper left panel in Fig.~\ref{Fig.fakeRH}). 
The ``fake'' RH was injected in the u-v data by means of the task UVMOD in AIPS and the resulting new data set was imaged using standard procedures. 
The brightness profile of the fake RH was modelled with a set of optically thin 
spheres with varying radius and flux density (Brunetti et al. 2007; Venturi et al. in prep).
The u-v data-set used for this procedure was chosen from VLA archive observations carried out in the D configuration and it was edited so as to mimic a typical NVSS field in terms of u-v coverage, sensitivity and observing beam.

In Fig.\ref{Fig.fakeRH} we report the results from the injection of fake RHs
with fixed radius, $R_H=500$ kpc and different total flux densities,
 and assuming z=0.1 (which is a mean value for the NVSS sample). 
From left top panel to right bottom panel we report
the results from injections of fake RHs 
with increasing total flux density. 
We find that diffuse emission is unambiguously 
imaged for RHs with flux density between 28 and 32 mJy.
At the assumed redshift this corresponds to 1.4 GHz luminosity of RHs 
of $P_{1.4} \sim7-8\cdot 10^{23}$ Watt/Hz, which corresponds to cluster X--ray
luminosity $L_X \sim 4\cdot 10^{44}$ erg/s (once the 
$P_{1.4}-L_X$ is assumed), in good agreement with our estimates
of the RH--selection effects in the NVSS in the most conservative case a) 
($R_H=500$ kpc, black line in Fig.~\ref{Fig.totsample}), which therefore 
will be adopted in the following of the paper.

\subsection{The GMRT sub-sample}

The case of the GMRT sub--sample is different from that of the NVSS 
sub--sample, because the observations carried out by our group 
(Venturi et al. 2007, Venturi et al in prep) are much deeper 
(1$\sigma\sim\,35-100~\mu$Jy beam$^{-1}$) and aimed at the detection of RHs.
This guarantees that the detection of extended diffuse emission at the
level expected from the $P_{1.4}-L_X$ correlation is not biased
by sensitivity limit of those observations.

Brunetti et al. (2007) derived solid upper limits to the 
radio powers of clusters (in the GMRT sample) without evidence 
of diffuse radio emission. These upper limits lie about one order 
of magnitude below the RH luminosities expected on the basis 
of the $P_{1.4}-L_X$ correlation.
This allows to conclude that the radio follow up performed for the
GMRT sub--sample does not introduce relevant biases to the detection of RHs in
the cluster sample.

\section{Results: fraction of galaxy clusters with radio halos}

Based on the results reported in the previous Sections, we define a 
new cluster sample which is made by the GMRT sub--sample (Sect.~2.2) 
and by all clusters of the XBACS--NVSS sub--sample (Sec.~2.1)
not affected by biases in the detection of RHs, namely all those  
above the curves in Fig.~\ref{Fig.totsample}.

In Fig.~\ref{Fig.totsample} the clusters hosting a giant RH are represented
with open red circles: it is clear that RHs most likely occur
in high X-ray luminosity clusters, and this is consistent with previous 
claims (GTF99).

To obtain a statistically meaningful estimate of this ``behaviour'' 
we focus on the most conservative sub--sample, which is made by all
clusters which lie above the upper curve in Fig.~\ref{Fig.totsample}; 
this curve is derived by adopting the case a) (Sec.3.1), 
that is supported by UVMOD simulations (Fig.~\ref{Fig.fakeRH}).
 
As already anticipated the REFLEX and eBCS samples are complete at 
the X-ray flux limit up to $z\approx0.3$, and Fig.~\ref{Fig.totsample} 
shows that incompleteness strongly affects 
the population of clusters with X-ray luminosity typical of the GMRT sub-sample
at $z\geq0.32$. Thus we limit our analysis to the redshift range $0.044-0.32$
and calculate the fraction of clusters with RHs in the population of low 
luminosity (LL) and high luminosity (HL) clusters. 
We chose $L_X=10^{44.9}$ erg/s as transition value between the LL and HL
samples. Such value provide good statistics in both samples
and ensure that the HL clusters cover the luminosity interval where the
occurrence of RHs seems to increase (Fig.~\ref{Fig.totsample}).
The LL sub-sample is thus composed 
by those clusters with X-ray luminosity above the black curve in Fig.~\ref{Fig.totsample} and below $10^{44.9}$ erg/s.    

\begin{figure}
\begin{center}
\includegraphics[width=7cm,height=7cm]{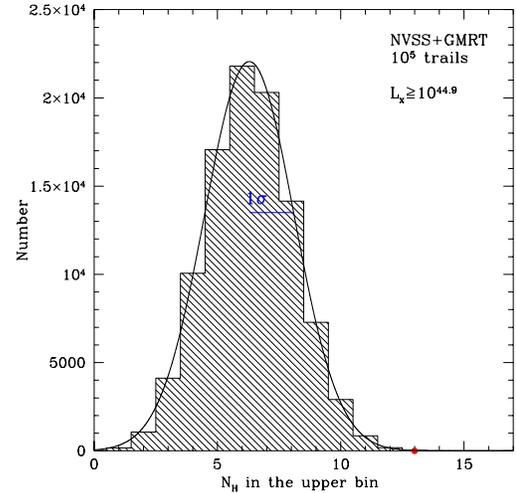}
\caption[]{Montecarlo realization of $10^5$ trials in the HL clusters
in the NVSS+GMRT sample.}
\label{Fig.montecarlo_44p9}
\end{center}
\end{figure}

We found that the fractions of clusters with RHs are
$f_{RH}\simeq 0.406\pm0.112$ and $f_{RH}\simeq 0.075\pm0.038$ in
the HL and LL sub-samples, respectively 
(1$\sigma$ Poissonian errors).

This is the most conservative case, indeed it is worth mentioning that 
by considering LL sub--samples defined by other selection--curves (case b) and c)), 
smaller values of $f_{RH}$ are obtained in the LL clusters:
$f_{RH}\simeq 0.063\pm0.026$ in the case b) (red dotted curve in Fig.~\ref{Fig.totsample}) and $f_{RH}\simeq 0.074\pm0.030$ in the case c) (2$sigma$, blue dot-dashed line in Fig.~\ref{Fig.totsample}). 

In all these cases there is a clear evidence of an increase of the
fraction of clusters with RHs with increasing the X-ray luminosity.
To test the strength of this result we consider the most conservative 
case and run Monte Carlo calculations. 
Specifically in this case the sample is composed by 85 clusters (32 are HL clusters
and 53 are LL clusters) with 17 RHs, of which 13 are hosted by HL clusters and 4 by LL clusters. 
We randomly assign 17 RHs among the 85 clusters of our sample
and count the number of RHs which fall into the HL sub-sample in our Monte Carlo trials. 
In Fig.~\ref{Fig.montecarlo_44p9} we report the distribution 
of the number of RHs in the HL sub-sample obtained after 
$10^5$ Monte Carlo trials. 
The distribution can be fitted with a Gaussian function with a central value 
of $\sim6.29$ and a standard deviation of $\sim1.81$.
This means that the observed value of 13 RHs in the HL clusters
is at $3.7\sigma$ from the value expected in the case 
of RH occurrence independent of cluster X--ray luminosity, 
and thus that the probability to have the observed fraction 
of clusters with RHs by chance is $\leq 0.2\cdot 10^{-3}$.
Here we have assumed the presence of a RH in the cluster A~2261.
This does not affect our result as the observed jump remains 
highly significant ($\sim 3.4\sigma$) even by assuming that A~2261 
does not have a RH.


This shows that our result is statistically significant and suggests, for the 
first time, that there is a {\it real} (\ie physical) increase of the occurrence 
of RHs with increasing the X-ray luminosity of the hosting clusters
in an X--ray selected sample.

\begin{figure}
\begin{center}
\includegraphics[width=7cm,height=7cm]{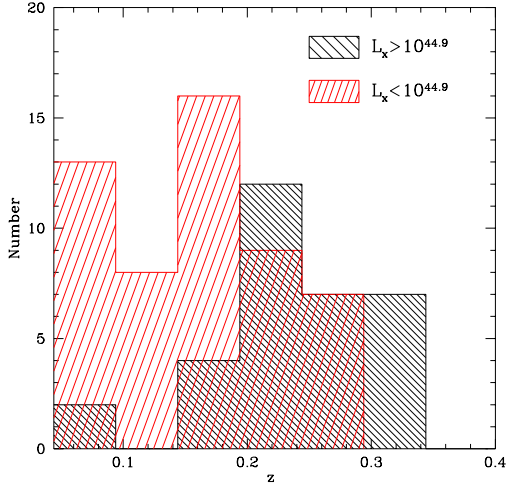}
\caption[]{Distribution of clusters as a function of redshift in the 
HL and LL sub-samples.}
\label{Fig.histo_z_44p9}
\end{center}
\end{figure}

In principle, because we are using X--ray selected cluster samples, 
it is not immediate to conclude if the increase of the fraction 
of clusters with RHs across the LL and HL clusters is purely 
driven by the X-ray luminosity of clusters or is due to possible 
difference in the redshift distribution of clusters in the two HL and LL
sub-samples.
In Fig.\ref{Fig.histo_z_44p9} we report the redshift distributions of 
clusters in the two sub-samples. 
HL clusters have typically larger redshift because they can be detected by X--ray surveys at larger cosmic distance.
More specifically, the mean redshift of LL clusters is $\sim 0.16$, while that of HL clusters is $\sim 0.23$.
However the difference in redshift distribution is relatively
small ($\sim 0.74$ Gyr in cosmic time) 
and thus it is very unlikely that the increase (a factor of $\sim5$) 
of the fraction of clusters with RHs observed in the two sub-sample 
is due to some cosmic evolution of the population
of RH sources. In addition, regardless of the origin of the
particles emitting RHs, it should be mentioned that the radiative life--time
of these particles decreases with redshift due to IC losses and thus,
in general, the fraction of clusters with RHs would be expected to decrease with 
cosmic look back time. Finally we note that the increase of the fraction of clusters with RH with cluster X-ray luminosity appears in Fig.~\ref{Fig.totsample}
also considering the GMRT sample alone, in which case clusters are essentially at the same redshift.
Thus, although it is true that both luminosity and redshift may influence 
the occurrence of RHs (and model calculations that account for both effects are discussed in Sec.~5), we believe that the X-ray luminosity plays the major role in
our cluster sample.
  
One possibility to further test this issue and to 
minimize the possible effect of the redshift on the observed
fraction of clusters with RHs is to consider the clusters of the 
XBAC-NVSS only which lie above the most conservative radio 
selection--curve in Fig.~\ref{Fig.totsample}; this 
sample is selected within $0.044<z<0.2$. 
In this case, however, it should be considered that 
the smaller cluster statistics and range  
of X--ray luminosities in the sample 
are expected to decrease the 
statistical significance of any possible trend observed
for the fraction of RH clusters with cluster properties.
Also in this case we derive the fraction of clusters with RHs in LL and HL 
clusters but to maintain a sufficient cluster statistics
in both the sub-samples we adopt an X-ray luminosity threshold of 
$\sim\,10^{44.8}$.

\begin{figure}
\begin{center}
\includegraphics[width=7cm,height=7cm]{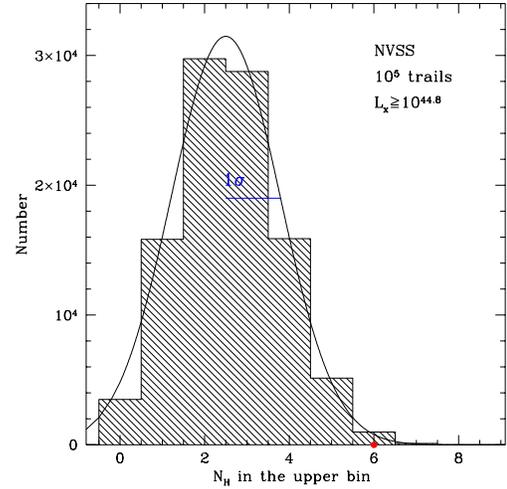}
\caption[]{Montecarlo realization of $10^5$ trials for the number 
of RHs in HL clusters in the NVSS-XBACs sample.}
\label{Fig.montecarlo_44p8}
\end{center}
\end{figure} 

We find $f_{RH}\simeq 0.43\pm0.17$ and $f_{RH}\simeq 0.088\pm0.051$ 
in the HL and LL clusters, respectively, 
which again suggests an increasing fraction of clusters 
with RHs with increasing X-ray luminosity. 
 
As in the previous case, in order to test the statistical strength of 
this results, we perform Monte Carlo calculations by 
assuming equal probability to have RHs in the clusters of the sample.
In this case we have a sample of 48 clusters (14 are HL clusters
and 34 are LL clusters) with 9 RHs, 6 of these RHs are
in the HL sub-sample. In Fig.\ref{Fig.montecarlo_44p8} we report 
the distribution of RHs in the HL sub-sample as counted after $10^5$ trials. 
The distribution is fitted with a Gaussian function which peaks 
at $\sim2.49$ and with a standard deviation of $\sim 1.29$, thus the observed 
value of 6 RHs is at $\sim2.7\sigma$ from that expected in the case
of equal probability to have RHs with cluster X-ray luminosity;
this means that the probability to obtain by chance the observed fraction of
clusters with RHs is $\leq 9\cdot 10^{-3}$.
 
Thus, although the significance of the result is reduced with respect 
to the case in which we consider the full NVSS+GMRT sample, the observed jump 
in the occurrence of RHs with the X-ray luminosity is still statistically
meaningful and further suggests that the effect is physically
driven by cluster X--ray luminosity (or mass).

\subsection{The effect of cooling core clusters on the observed statistics}

\begin{figure*}
\begin{center}
\includegraphics[width=0.35\textwidth]{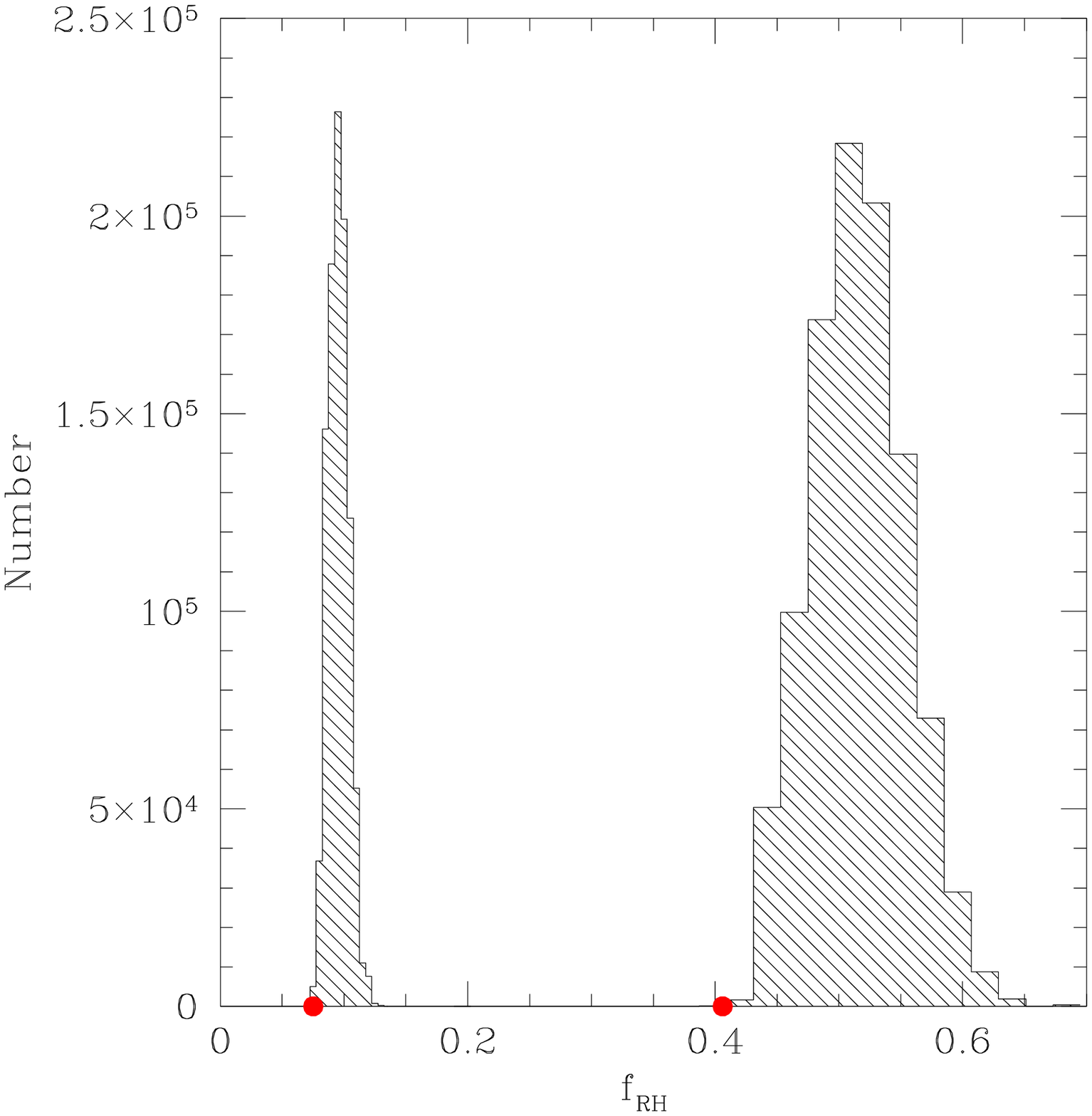}
\includegraphics[width=0.35\textwidth]{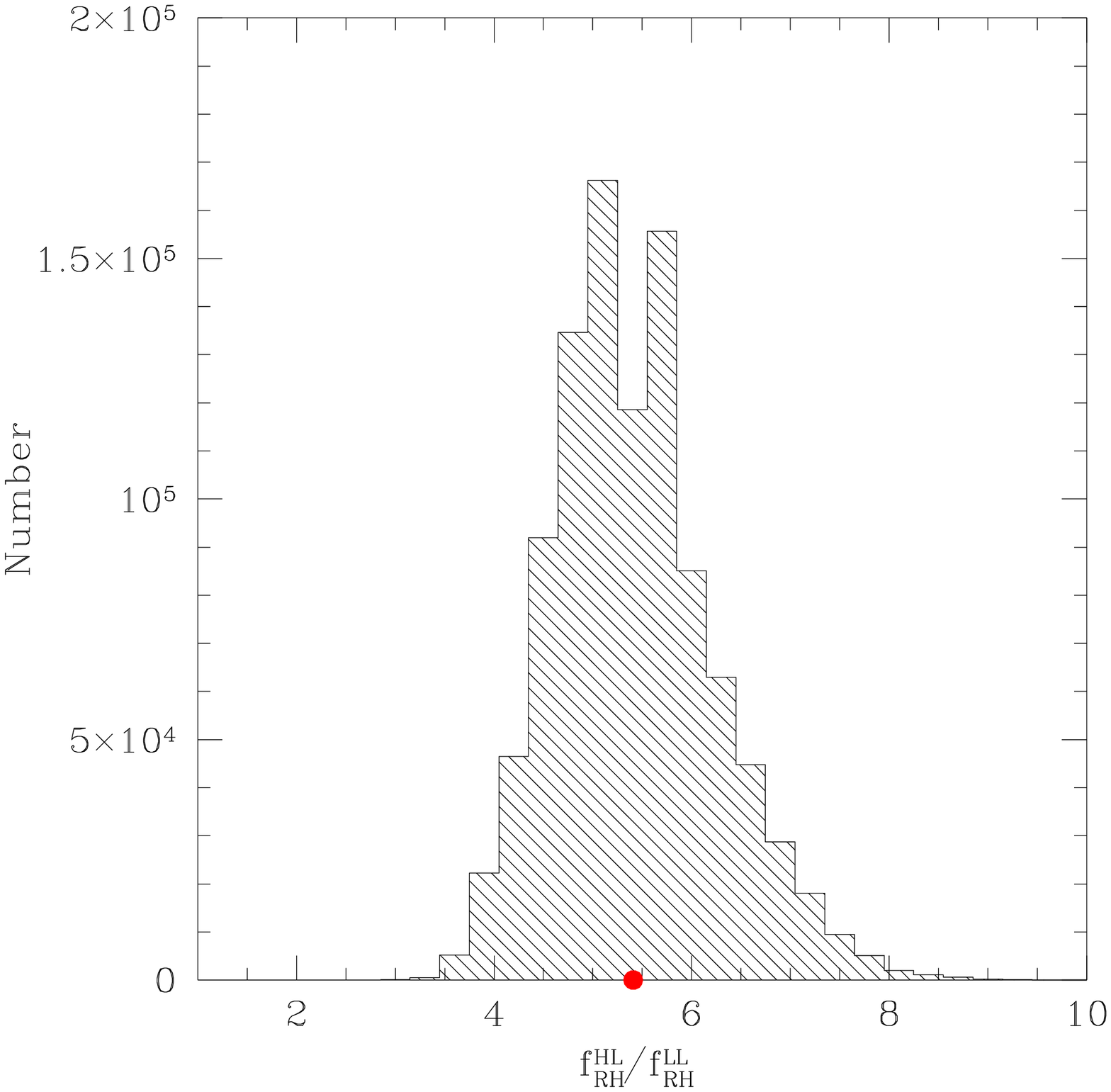}
\caption[]{a) Distributions of the fractions of clusters with RH in LL (left histogram) and HL (right histogram) sub-sample, obtained after $10^{6}$ Montecarlo trials. The red filled circles rapresent the corresponding value of $f_{RH}$ obtained without CC correction. b) Distribution of the ratio between the fraction of clusters with RH in HL and LL clusters obtained from $10^6$ Montecarlo trials, the red filled circle being the value obtained without CC correction.}
\label{Fig.MC_cool}
\end{center}
\end{figure*}

It should be mentioned that the $P_{1.4}-L_X$ correlation we used
in Sec.~3 to derive the capability of the NVSS in catching RHs, is found for RH clusters which do not have a CC.
In principle, the X-ray luminosity of these clusters should be corrected for the contribution of the CC when using the $P_{1.4}-L_X$ correlation. Due to the lack of an adequate X-ray follow up of
the clusters in our sample, it is impossible to identify those affected by a CC.
However the fraction of CC clusters is found at a level of $\sim30\%$ for BCS clusters with $L_X\geq4\cdot 10^{44}$ at $z\sim 0.15-0.4$ (Bauer et al. 2005) and this is unlikely to affect our main results.
Indeed once corrected from the CC contributions some HL clusters may become 
LL clusters and some LL clusters may fall below the X-ray luminosity
threshold and leave the selected sample. Because the LL and HL clusters are comparable
in number, the occurrence of RH in both HL and LL clusters is expected to
remain unchanged (although the fraction of RH in HL clusters may slightly increase). 
In order to better quantify this effect we work out Montecarlo simulations. 
We assume that $30\%$ of galaxy clusters of our sample are CCs and  
randomly assign CCs among clusters without RHs. 
Once the CC is assigned to a cluster
we apply a correction to its X-ray luminosity which is reduced by a factor randomly chosen between 1.4 and 2.5 (consistent with recent observations,\eg Zhang et al. 2006, 2007; Chen et al. 2007) and re-evaluate the statistics of RH in the ``corrected'' 
LL and HL samples. In Fig.~\ref{Fig.MC_cool} a) we report the distributions of the fractions of clusters with RH in the ``corrected'' sample, LL (left distribution) 
and HL (right distribution), obtained after having applied our procedure in $10^6$ Montecarlo trials. 
The fractions of clusters with RHs in both LL and HL clusters only slightly increase with respect to the previous statistical calculation (Sect.4). 
In Fig.~\ref{Fig.MC_cool} b) we report the distribution of the ratio between the fraction of clusters with a RH in the ``corrected'' HL and LL samples obtained from $10^6$ Montecarlo trials: it is clear that the bulk of the values is consistent with that from the previous statistical analysis. 
These results, based on viable assumptions, show that the main finding of our work, the increase of the RH occurrence with the X-ray luminosity of galaxy clusters, 
might not be affected by the possible presence of CC clusters in the sample.

\section{A comparison between model expectations and observations}

\begin{figure}
\includegraphics[width=0.5\textwidth]{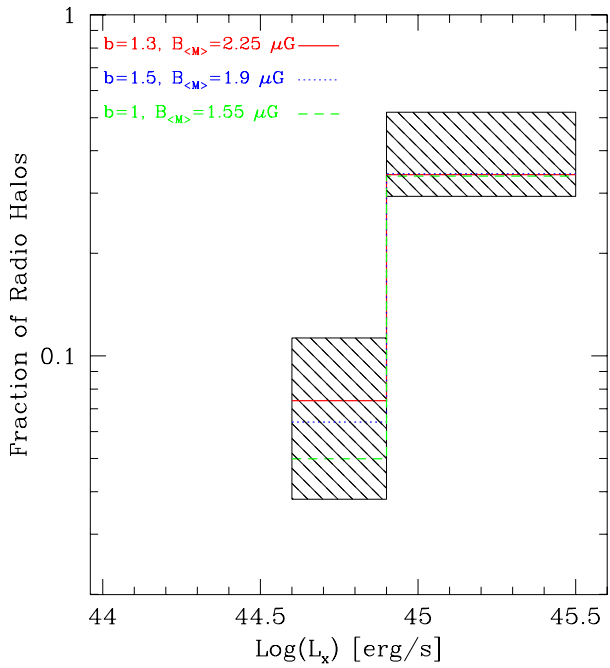}
\caption[]{Expected fraction of clusters with RHs in the HL and LL sub-samples
(different colored lines) vs the observed occurrence of RH with 1$\sigma$ errors 
(shadowed regions).}
\label{Fig.theo_obs_Lx}
\end{figure} 

In this Section we provide a basic comparison between present observations
and the expectations of the re-acceleration model in its simplest form.
In the previous Section we showed that the fraction of clusters
with RHs increases with the cluster X-ray luminosity and 
this provides an important piece of information that should be 
explained by the theoretical pictures for the origin of RHs.
CB05 modelled the statistical properties of RHs in the framework of the 
merger--induced {\it in situ} particle acceleration scenario. 
By adopting the semi--analytic Press \& Schechter (1974) theory 
to follow the cosmic evolution and formation of a large synthetic 
population of galaxy clusters, it was assumed that the energy injected 
in the form of magnetosonic waves during merging events in clusters 
is a fraction, $\eta_t$, of the $PdV$ work done by the infalling 
subclusters in passing through the most massive one. 
Then the processes of stochastic acceleration of the relativistic
electrons by these waves, and the ensuing synchrotron emission properties,
have been calculated under the assumption of a constant magnetic 
field intensity averaged within a 1 Mpc$^3$ volume. 
CBS06 have extended this analysis, by including a scaling of 
the magnetic field strength with cluster mass in the form $B \propto M_v^b$. 
They showed that the observed correlations between the synchrotron radio power 
of a sample of 17 RHs and the X-ray properties of the hosting clusters 
can be reproduced in the framework of the re-acceleration model for 
$b \gtsim 0.5$ and typical $\mu$G strengths of the average $B$ intensity.
Those values provide a working framework in the range of ($B,b$) model parameters
under which the statistical properties of RHs, namely the occurrence 
of RHs with cluster mass and z, the luminosity functions and number counts, 
have been extensively calculated.

An important finding of this work was that generally the probability 
to form giant RHs increases with cluster mass, which is a unique 
expectation of the re--acceleration scenario.
In particular an abrupt increase of the fraction of clusters with RHs is expected across cluster masses of $\sim 2\cdot 10^{15}\, M_{\odot}$, since the turbulent energy injected during cluster mergers scales with the cluster thermal 
energy (which roughly scales as $\sim M^{5/3}$), and turbulence 
in more massive clusters is injected on larger volumes (see CB05).

Specifically, the most important goal of this Section is to test whether this re-acceleration scenario can potentially match the increase of the fraction of clusters with RHs as derived from our cluster sample. We therefore adopt the same approach
outlined in CB05 and CBS06 and compare model expectations with the results obtained
in the previous section (case NVSS+GMRT samples).

In order to have a prompt comparison between model expectations 
and the data we calculate the expected fraction of clusters with RHs in different redshift bins by assuming the same redshift distribution of galaxy clusters as in 
Fig.~\ref{Fig.histo_z_44p9}.
It should be mentioned that the model use virial masses ($M_v$)
and thus to compare our expectations with observations we use the observed 
$L_X-M_v$ correlation (Eq.~7 in CBS06): $L_X\propto M_v^{1.47}$.
This correlation is used to convert the range of X--ray luminosity of
LL and HL clusters into mass ranges in which to compute the expected 
fraction of clusters with RHs.

In Fig.~\ref{Fig.theo_obs_Lx} we report the derived $f_{RH}$ for different values of the model parameters (different lines) as a function of the X-ray luminosity overlaid on the observed values of $f_{RH}$ (shadowed region). The model parameters are the magnetic field intensity, $B_{<M>}$, of a cluster with virial mass $<M>=1.6\cdot 10^{15}\,M_{\odot}$ and the slope $b$ of the scaling of $B$ with $M_v$ (see CBS06 for details). In all the three cases a value of $\eta_t\simeq0.18$ is adopted:
this value falls in the range of $\eta_t$ that has been shown to 
allow the re-acceleration model, in the form developed by CB05 and CBS06, to be consistent with the production of RHs in $\approx 1/3$ of massive galaxy clusters.
We find that the expected behaviour of the occurrence of RHs with cluster mass is consistent with the observed one for all the considered values of parameters.

It should be mentioned that the use of the $L_X-M_v$ correlation 
implies that $L_X$ reported in Fig.\ref{Fig.theo_obs_Lx} is
a good proxy for the mass.
This $L_X-M_v$ correlation is obtained in CBS06 by using the HIFLUGCS 
cluster sample (Reiprich \& B\"ohringer 2002) which contains CC and non 
CC clusters and which is characterized by a relatively large
scatter in the mass, \iee $\approx 45\%$ (Reiprich \& B\"ohringer 2002).
Very recent results on the HIFLUGCS sample showed that 
most of this scatter is due to the mixing of CC and non CC clusters
in the sample, with CC and non CC clusters following 
correlations with compatible slopes (within 1 $\sigma$) but different normalizations\footnote{This is shown for both the $L_X-M_{500}$ and
$L_X-M_{200}$ correlations.} (Chen et al. 2007; Reiprich \& Hudson 2006).

\noindent Thus the use of the $L_X-M_v$ correlation of CBS06
implies that the masses of non CC clusters and CC clusters with a 
fixed $L_X$ would be statistically under--estimated and over--estimated, 
respectively, and this would provide a compromise in the
calculation of the masses spanned by the LL and HL clusters.
A more detailed comparison would require the 
knowledge of the virial masses for all the clusters of our sample,
which is however not available at the moment. Nevertheless this is 
unlikely to affect our results significantly. Indeed one simple check 
is to use the $f_{RH}$ in Fig.~\ref{Fig.MC_cool}
(\ie obtained from Montecarlo simulations after statistical 
correction of the $L_X$ of CC clusters) and to apply a systematic 
positive correction of $\approx 30$ \% to the mass of LL and HL clusters
with respect to the case of the $L_X-M_v$ correlation of CBS06
\footnote{This correction has been estimated from the $L_X-M_{200}$
correlation for non CC clusters in Fig.~2 of Reiprich \& Hudson 2006.}.
In this case we checked that the jump in $f_{RH}$ and the values of
$f_{RH}$ themselves in Fig.~\ref{Fig.MC_cool} can be still well reconciled
with the model expectations\footnote{We note that in this case, because
the cluster masses (energetics) are slightly larger, the model formally
reproduces $f_{RH}$ for a slightly smaller value of $\eta_t$, 
$\eta_t\approx0.16$.}.

\begin{figure*}
\includegraphics[width=0.45\textwidth]{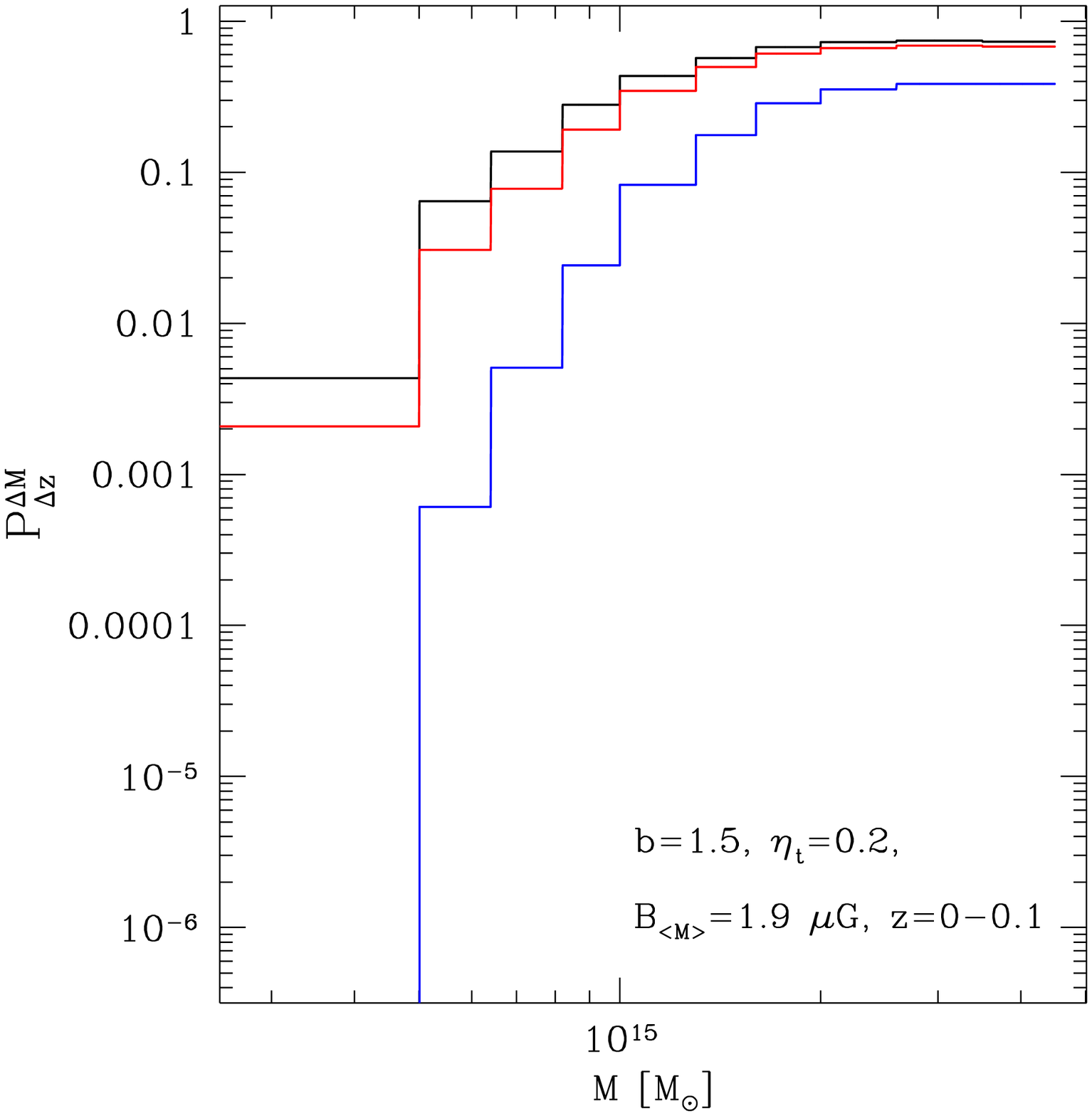}
\includegraphics[width=0.45\textwidth]{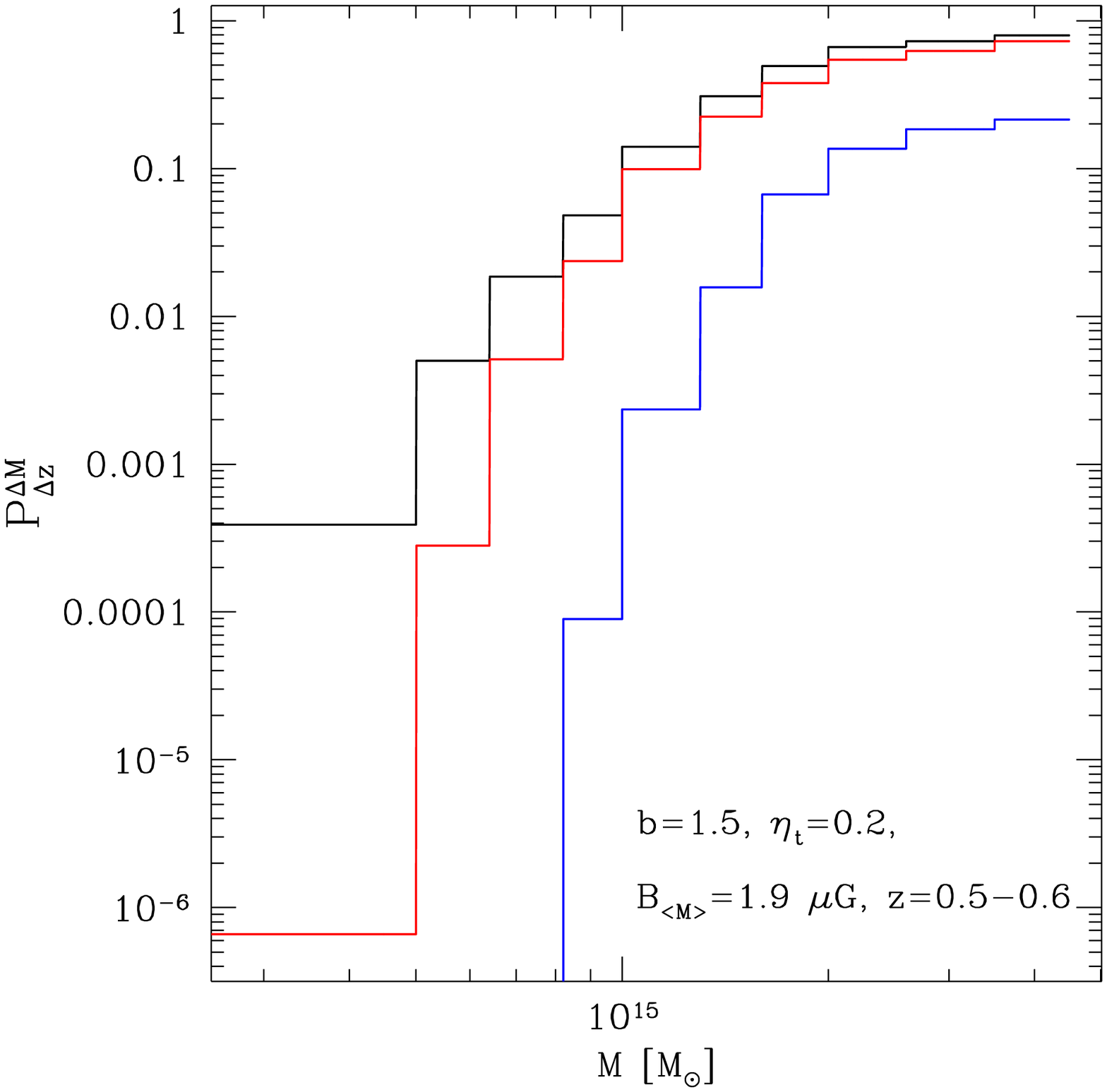}
\caption[]{Expected fraction of clusters with RHs at three
different radio frequencies: 1.4 GHz (blue lines), 240 MHz (red lines) and
150 MHz (black lines) in the redshift bin $z\sim 0-0.1$ (left panel) and 
$z\sim 0.5-0.6$ (right panel). Calculations have been performed assuming:
$b=1.5$, $B_{<M>}=1.9$ $\mu$G and $\eta_t=0.18$.}
\label{Fig.theo_lofar}
\end{figure*} 

\section{Radio Halos in future low-frequency radio surveys}

In the previous Section we showed that the 
re-acceleration scenario may explain the observed increase of the fraction of clusters with RHs with cluster X-ray luminosity (mass).
On the other hand, from Fig.\ref{Fig.theo_obs_Lx} it is evident 
that with the present surveys it is difficult to make a more quantitative statement, \ie the poor statistics does not allow to disentangle among different configurations of the model parameters.

In a few years LOFAR and LWA will survey galaxy clusters with unprecedented	
sensitivity, thus increasing the observational statistics and hopefully providing 
a much clear situation. On the other hand 
these radio telescopes will observe at frequencies $\approx 10$ times lower than the present facilities and this could make the interpretation of the observations more tricky.
Indeed another natural expectation of the re-acceleration scenario 
is that the fraction of clusters with RHs depends on the observing frequency. 
In particular, it is expected that the number of RHs should increase 
at low radio frequencies, $\approx 100$ MHz, since low energy electrons (with Lorentz factor $\gamma\approx 3000$) are able to radiate synchrotron emission at these frequencies 
and may be re-accelerated also in less energetic merger events (CBS06). For energetic reasons it is also expected that this increase should be more pronounced in the case of less massive clusters which thus should host a sizable fraction of the population of RHs emitting at low radio frequencies.

The new generation of low frequency interferometers, LOFAR and LWA, 
represents therefore a unique possibility to discover a {\it new class
of Radio Halos} predicted on the basis of the re-acceleration scenario which may emerges {\it only} at low radio frequencies.
In particular, it is expected (CBS06) that the number of {\it low frequency RHs} should be $\approx10$ times that of ``classical'' RHs (those emitting at $\approx$GHz frequencies), thus the discovery of an excess of RHs at low frequencies 
with respect to the extrapolation of the number counts of present RHs  
would be a prompt confirmation of this scenario.

In Fig.\ref{Fig.theo_lofar} we report some preliminary calculations (based on CBS06, Sect.~7) of the probability to have RHs at 
different radio frequencies ($1.4$ GHz, $240$ MHz and $150$ MHz, see figure caption).
We found a large increase of the expected fraction of
clusters with RHs at low radio frequencies ($150$-$240$ MHz); this
increase is even more striking for less massive clusters,
\ie $M<10^{15}\,M_{\odot}$. Indeed if the fraction of clusters with RHs in the redshift bin $0-0.1$ increases between $1400-150$ MHz by a factor of $\sim 2$ for $M\sim 2-3\cdot 10^{15}\,M_{\odot}$, this increase is a factor of $\gtsim 10$ for 
$M \ltsim 10^{15}\, M_{\odot}$. Furthermore, at higher redshifts,
$z\sim 0.5-0.6$, this increase is even larger, a factor of $\sim 4$ and 
$\gtsim 10^2$ in the case of $M\sim 2-3\cdot 10^{15}\,M_{\odot}$ and $M \ltsim 10^{15}\, M_{\odot}$, respectively. From these results, we reach the important conclusion that the increase of the fraction of clusters with RHs with increasing cluster mass (or X-ray luminosity) should become less striking at low radio frequencies.
In particular, considering the mass range in which present radio surveys
have observed RHs, \ie $10^{15} M_{\odot}\ltsim M\ltsim 4\cdot 10^{15}M_{\odot}$, and considering two population of clusters with a separation mass $\sim 2\cdot 10^{15}\,M_{\odot}$,
the jump in the expected fraction is a factor of $\sim 4$ at 1.4 GHz 
and is reduced to a factor of $\sim2$ at 150 MHz.
Thus, in general, LOFAR and LWA surveys are expected to find a smaller increase of the 
fraction of clusters with RHs with the cluster mass with respect to present surveys.

\section{Conclusions}

The statistical properties of RHs are an important piece of information for the present understanding of non-thermal phenomena in galaxy clusters:
do all clusters have a RH? Does the fraction of clusters with 
RHs depend on cluster mass and redshift? 

In the past a few seminal attempts were made to address
these points (Giovannini et al. 1999; Kempner \& Sarazin 2001), 
however it was not clear how much those results were affected by the sensitivity 
limit of the adopted radio surveys. In particular it was unclear 
whether the absence of RHs in low X-ray luminosity clusters could be ascribed 
to the steepness of the radio power--X-ray luminosity correlation of RHs combined 
with the sensitivity limit of the adopted 
surveys (Kempner \& Sarazin 2001). 

In this work we presented an unbiased statistical analysis of RHs in a large sample of clusters with radio information. We combined the sub-sample of XBACs clusters at $z\leq 0.2$, checked for the presence of diffuse radio emission in the NVSS by GTF99, with two sub-samples of clusters in the redshift range $0.2-0.4$ (extracted from the REFLEX and eBCS catalogs) inspected at 610 MHz and at 1.4 GHz with deep GMRT and VLA observations, the GMRT sub-sample (Venturi et al. 2007; Venturi et al. in prep.).

Brunetti et al. (2007) showed that the bulk of clusters of the GMRT sample does not show any hint of Mpc scale diffuse radio emission at their centre at the level of presently known RHs. 
For the clusters without extended radio emission upper limits to their radio power 
were obtained: they are about one order of magnitude below the values expected on the basis of the radio power-X-ray luminosity correlation, and this allowed to
conclude that such correlation is {\it real}, \iee not driven by observational biases at least for $L_{X}\ge 5\cdot 10^{44}$ erg/s (the lower X-ray luminosity of the GMRT sample).

Assuming that RHs follow this correlation, in this paper we have obtained the minimum X-ray luminosity of clusters which may host a RH detectable in the NVSS as a function of z. We followed a number of approaches based on the brightness profiles of well
studied RHs. In the most conservative case (case a), Sect.~3.1) we successfully test this approach by means of injection of
{\it fake} RHs in a NVSS-like UV dataset.
Based on these results for the NVSS and on those of Brunetti et al. (2007)
for the GMRT sample, we were able to select a sample of clusters (NVSS+GMRT) which is not affected by observational biases and which is suitable 
for a statistical analysis. 

The main result of our analysis of this sample is that we find an increase of the fraction of clusters with RHs with the cluster X-ray luminosity.
More specifically, we find that the percentage of clusters at $z\approx 0.044-0.32$
with $4\cdot 10^{44}\, erg/s\,\ltsim L_X\ltsim 8\cdot 10^{44}\,erg/s$ hosting RHs
is $7.5\pm3.8\%$, while that of clusters with $L_X\ge 8\cdot 10^{44}$ erg/s
is $40.6\pm 11.2\%$.
We tested the significance of this result by means of Montecarlo trials which
allow us to conclude that the observed jump is {\it real}, with a significance of $3.7\sigma$. In Sec.~4.1 we also showed that this result is not 
appreciably affected by the possible presence of CC clusters in the sample.

We showed that the increase of the fraction of clusters with RHs
can be understood in the framework of the re-acceleration scenario.
This increase is a natural expectation of this scenario since a more efficient particle acceleration is triggered in massive (and X-ray luminous) clusters 
during merger events.

Since this scenario is at present in line with several observations, 
it is extremely intriguing, in view of forthcoming low frequency interferometers, such as LOFAR and LWA, to investigate the model expectations at low radio frequencies, \iee few hundreds of MHz.
In Sec.~6 we showed that this model expects that the population of low frequency RHs is $\approx10$ times larger than that of ``classical'' RHs (those emitting at $\approx$ GHz frequencies) and we showed that the increase of the probability to have RHs emitting at low radio frequencies is maximized in the case of clusters with lower masses
($M\ltsim 10^{15}\,M_{\odot}$). This implies that the increase of the fraction of
clusters with RHs with cluster mass as measured by LOFAR and LWA surveys is expected to be less pronounced than that observed by present surveys.

\begin{acknowledgements}
This work is partially supported by MIUR, ASI and INAF under grants PRIN2004, PRIN2005, PRIN-INAF2005 and ASI-INAF I/088/06/0. R.C. thanks F. Vazza for useful discussions. We
acknowledge the anonymous referee for useful comments.
\end{acknowledgements}

\end{document}